\documentclass[final,1p,times]{elsarticle} 

\usepackage{graphicx}
\usepackage{amssymb} 
\usepackage{amsthm} 
\usepackage{lineno}

\def\sNN{\mbox{$\sqrt{s_{_{NN}}}$}}
\newcommand{\prl} {Phys. Rev. Lett.}
\newcommand{\prc} {Phys. Rev. C}
\newcommand{\prd} {Phys. Rev. D}
\newcommand{\npa} {Nucl. Phys. A}
\newcommand{\plb} {Phys. Lett. B}
\newcommand{ \be }{\begin{equation}}
\newcommand{ \ee }{\end{equation}}
\newcommand{ \bea }{\begin{eqnarray}}
\newcommand{ \eea }{\end{eqnarray}}

\journal{Nuclear Physics A} 

\begin{document}

\begin{frontmatter} 

\title{Search for Chiral Magnetic Effects in High-Energy Nuclear Collisions}

\author{Gang Wang (for the STAR\fnref{col1} Collaboration)}
\fntext[col1] {A list of members of the STAR Collaboration and acknowledgements can be found at the end of this issue.}
\address{Department of Physics and Astronomy, University of California, Los Angeles, CA, USA}


\begin{abstract} 
We present measurements of pion elliptic flow ($v_2$) in Au+Au
collisions at $\sNN =$ 200, 62.4, 39, 27 and 19.6~GeV,
as a function of event-by-event charge asymmetry ($A_{\pm}$),
based on data from the STAR experiment at RHIC.
We find that $\pi^-$ ($\pi^+$) elliptic flow linearly increases (decreases) with
charge asymmetry for most centrality bins and for all the beam energies under study.
The slope parameter ($r$) from $v_2(A_\pm)$ difference between $\pi^-$ and $\pi^+$
shows a centrality dependency similar to calculations of the Chiral Magnetic Wave.
The measurements of charge separation with respect to the reaction plane in search of
Local Parity Violation and the Chiral Magnetic Effect are also presented
for Au+Au collisions at $\sNN =$ 200, 62.4, 39, 27, 19.6, 11.5 and 7.7~GeV,
and for U+U collisions at 193~GeV.
\end{abstract} 

\end{frontmatter} 


\section{Introduction}

In heavy ion collisions at the Relativistic Heavy Ion Collider (RHIC) and the Large Hadron Collider (LHC),
energetic spectator protons produce a strong magnetic field peaking around $eB_y \approx m^2_\pi$~\cite{Kharzeev}.
The interplay between the magnetic field and the quark-gluon matter created in the collisions
is characterized by two phenomena: the Chiral Magnetic Effect (CME) and the Chiral Separation Effect (CSE).
The CME is the phenomenon of electric charge separation along the axis of the magnetic field
in the presence of a finite axial chemical potential
(e.g. chiral potential due to fluctuating topological charge)~\cite{Kharzeev,Kharzeev2,Kharzeev3,Kharzeev4,Kharzeev5}.
STAR~\cite{STAR_LPV1,STAR_LPV2} and PHENIX~\cite{PHENIX_LPV1,PHENIX_LPV2} collaborations at RHIC
have reported experimental observations of charge asymmetry fluctuations possibly providing an evidence for CME.
This interpretation is still under intense discussion, see e.g.~\cite{dis1,dis2} and references therein.
The Chiral Separation Effect (CSE) refers to the separation of chiral charge along the axis of
the magnetic field at finite density of vector charge (e.g. electric charge)~\cite{CSE1,CSE2}.

In a chirally symmetric phase, the CME and CSE effects form a collective excitation, Chiral
Magnetic Wave (CMW), a long wavelength hydrodynamic mode of chiral charge densities~\cite{CMW,CMW2}.
The CMW manifests itself in a finite electric quadrupole moment of the collision system,
where the ``poles" (``equator") of the produced fireball acquire additional positive (negative) charge~\cite{CMW}.
This effect, if realized, will be reflected in the measurement of charge-dependent elliptic flow.
Elliptic flow, characterized by a {\it second}-order harmonic in the particle azimuthal distribution ($\phi$)
and quantified by the Fourier coefficient $v_2$~\cite{Methods}, refers to the collective motion of particles
with respect to the reaction plane ($\psi_{\rm RP}$):
\be
v_2 = \langle \cos[2(\phi - \psi_{\rm RP})]  \rangle.
\ee
Taking pions as an example, on top of the baseline $v_2^{\rm base}(\pi^\pm)$, a CMW will lead to~\cite{CMW}
\be
v_2(\pi^\pm) = v_2^{\rm base}(\pi^\pm) \mp (\frac{q_e}{\bar \rho_e})A_{\pm},
\ee
where $q_e$, ${\bar \rho_e}$ and $A_\pm = ({\bar N_+} - {\bar N_-})/({\bar N_+} + {\bar N_-})$ are
the quadrupole moment, the net charge density and the charge asymmetry of the collision system, respectively.
As $\langle A_\pm \rangle$ is always positive, $A_\pm$-integrated $v_2$ of $\pi^-$ ($\pi^+$)
should be above (below) the baseline due to CMW.
However, the baseline $v_2$ may be different for $\pi^+$ and $\pi^-$ in the first place
because of several other possible physics mechanisms~\cite{other1,other2,other3},
so it is less ambiguous to study CMW via the $A_\pm$ dependency of pion $v_2$ than $A_\pm$-integrated $v_2$.

In Sec. 2, we present $A_\pm$-differential measurements of pion $v_2$ for Au+Au collisions
at $\sqrt{s_{NN}}=$200, 62.4, 39, 27 and 19.6~GeV.
We find that pion $v_2$ exhibits a linear dependence on $A_\pm$,
with positive (negative) slopes for $\pi^-$ ($\pi^+$).
The slope difference between $\pi^-$ and $\pi^+$ is studied as a function of collision centrality.
In Sec. 3, the measurements of charge separation correlator are used to search for
CME and Local Parity Violating (LPV) effects.

\section{Pion $v_2({A_{\pm}})$}

\begin{figure}[htbp]
\begin{center}
 \includegraphics[width=1.0\textwidth]{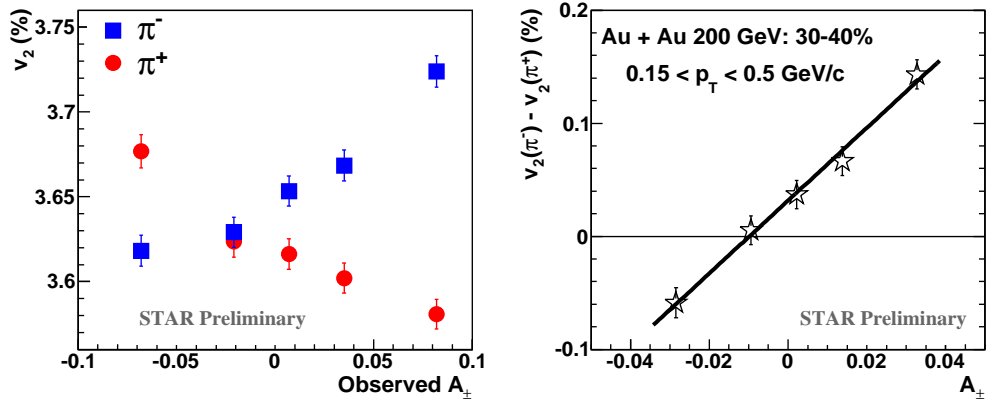}
\end{center}
\caption{(Color online) The example of $30$-$40\%$ Au+Au collisions at 200 GeV~\cite{Hongwei}. 
(Left) Pion $v_2\{2\}$ as a function of observed charge asymmetry. 
(Right) $v_2$ difference between $\pi^-$ and $\pi^+$ as a function of charge asymmetry with the tracking efficiency correction.
The errors are statistical only.}
\label{fig:1}
\end{figure}

Charged particle tracks were reconstructed in STAR TPC~\cite{TPC-NIM}, with pseudorapidity cut $|\eta|<1$.
The centrality definition and track quality cuts are the same as in Ref.~\cite{Flow200GeV}, unless otherwise specified.
This study is based on Au+Au samples of 200M events at 200 GeV from RHIC year 2010, 60M
at 62.4 GeV (2010), 100M at 39 GeV (2010), 40M at 27 GeV (2011) and 20M at 19.6 GeV (2011).
All were obtained with a minimum-bias trigger.
Only events within 40 cm of the center of the detector were selected for this analysis.
In the calculation of charge asymmetry, (anti)protons with transverse momentum $p_T<0.4$ GeV/$c$ were excluded
to reject beam pipe protons.
A distance of the closest approach (dca) cut ($< 1$ cm) was also applied to reduce the
number of weak decay tracks or secondary interactions.
To select pions, we eliminate charged particles $2\sigma$ away from the expected TPC energy loss for pions.

Elliptic flow measurements were carried out with the two-particle cumulant method 
$v_2\{2\}$~\cite{Methods,Flow200GeV} for 200 and 62.4 GeV,
and $v_2$\{$\eta$ sub\} approach for the rest beam energies,
where two subevents consist of charged particles with $\eta>0.3$ and $\eta<-0.3$, respectively.
Pions at positive (negative) $\eta$ are then correlated with the subevent at negative (positive) $\eta$
to calculate $v_2$.
The $\eta$ gap of 0.3 unit suppresses short-range correlations such as Bose-Einstein interference and
Coulomb final-state interactions~\cite{Flow200GeV}.
The $\eta$ gap was also used in the $v_2\{2\}$ analysis in a similar way.
To focus on the soft physics regime, only pions with $0.15 < p_T < 0.5$ GeV/$c$ were used to
calculate the $p_T$-integrated $v_2$.
Taking $30$-$40\%$ 200 GeV Au+Au as an example~\cite{Hongwei}, we show pion $v_2$ as a function of observed charge asymmetry
in the left panel of Fig.~\ref{fig:1}. $\pi^-$ $v_2$ increases with the observed $A_{\pm}$ while
$\pi^+$ $v_2$ decreases with a similar magnitude of the slope. After the tracking efficiency correction
for the charge asymmetry, the $v_2$ difference between $\pi^-$ and $\pi^+$ is fit with a straight line
in the right panel. The slope parameter $r$, or $2q_e/{\bar \rho_e}$ from Eq.~2, is positive and qualitatively
consistent with the expectation of the CMW picture. The intercept of the linear fit is non-zero,
indicating the baseline $v_2$ for $\pi^-$ and $\pi^+$ are different.

\begin{figure}[htbp]
\begin{center}
 \includegraphics[width=1.0\textwidth]{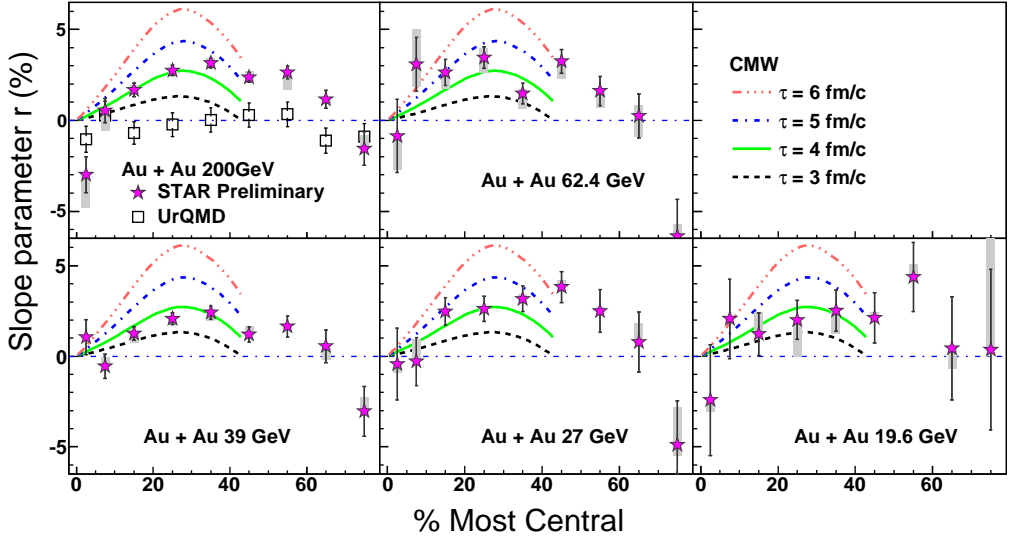}
\end{center}
\caption{(Color online) The slope parameter $r$, supposedly $2q_e/{\bar \rho_e}$, as a function of centrality. 
For comparison, we also show the UrQMD~\cite{UrQMD} simulation for 200 GeV Au+Au, and the calculations with
CMW~\cite{CMW} with different duration times.
The grey band represents the systematic uncertainty due to varied dca cuts and the tracking efficiency.}
\label{fig:2}
\end{figure}

In both $v_2\{2\}$ and $v_2$\{$\eta$ sub\}, there are correlations not related to the reaction plane,
and not suppressed by the $\eta$ gap, for example due to back-to-back jets.
They are largely canceled out in the $v_2$ difference between $\pi^-$ and $\pi^+$.
Correlations between daughters of weak decays like $\Lambda / {\overline \Lambda}$ may not be canceled,
and may contribute to the intercept of $v_2(A_{\pm})$ difference.
However, for neutral particles like $\Lambda$, the decay daughters on average will not contribute
to the numerator of observed $A_{\pm}$, so they will not create a correlation between $v_2$ difference
and $A_{\pm}$ out of nothing. In other words, this effect will not change the slope parameter from zero to finite.
The denominator of observed $A_{\pm}$ will be increased due to this effect,
and thus the observed slope parameter will be increased by a scale factor, related to the $\Lambda/\pi$ ratio,
which requires further systematic study.

We follow the same procedure as above to retrieve the slope parameter $r$ for all centrality bins
and all the collision systems under study. The results are shown in Fig.~\ref{fig:2}, together with
the simulation calculations with UrQMD event generator~\cite{UrQMD} and with the theoretical 
calculations with the CMW effect with different duration times for the magnetic field~\cite{CMW}.
For most data points, the slopes are positive and reach a maximum in mid-celtral/mid-peripheral collisions.
The slopes extracted from UrQMD events are consistent with zero for $15$-$60\%$ collisions, where the signal from the real data is prominent.
On the other hand, the CMW calculations demonstrate a similar centrality dependency of the slope parameter,
though quantitative comparison between data and theory requires further works on both sides
to match the kinematic regions of the analyses.

\section{Charge separation with respect to the reaction plane }

\begin{figure}[htbp]
\begin{center}
 \includegraphics[width=1.0\textwidth]{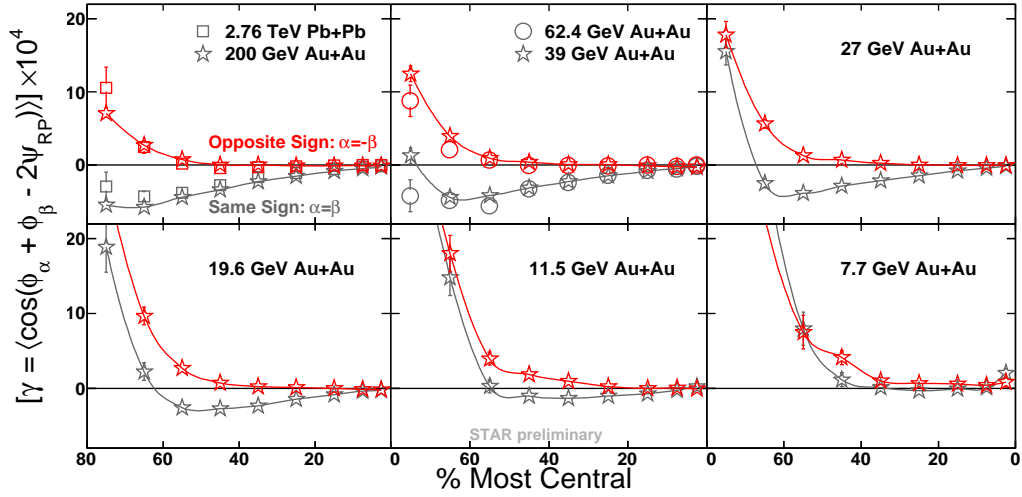}
\end{center}
\caption{(Color online) The three-point correlator, $\gamma$, as a function of centrality for Au+Au collisions 
from 200 GeV to 7.7 GeV~\cite{Dhevan}.
For comparison, we also show the results for Pb+Pb collisions at 2.76 TeV~\cite{ALICE}.
The errors are statistical only.}
\label{fig:3}
\end{figure}

The concept of Local Parity ($\cal P$) Violation (LPV) in high-energy heavy ion collisions
was brought up by Lee {\it et al.}~\cite{Lee, Wick, Morley} and elaborated by Kharzeev {\it et al.}~\cite{Dima}.
In non-central collisions such a $\cal P$-odd domain can manifest itself via preferential
same charge particle emission for particles moving along the system's angular momentum,
due to the Chiral Magnetic Effect~\cite{Kharzeev2,Kharzeev3}.
To study this effect, a three-point mixed harmonics azimuthal correlator was proposed~\cite{Sergei2}:
\be
\gamma = \langle \cos(\phi_{\alpha} + \phi_{\beta} - 2\psi_{\rm RP}) \rangle,
\label{eq:eq3}
\ee
where $\alpha$ and $\beta$ denote the particle type: $\alpha$, $\beta = +$, $-$.
The observable $\gamma$ is {$\cal P$}-even, but sensitive to the fluctuation of charge separation.
STAR measurements of the correlator were reported for Au+Au and Cu+Cu collisions at 200 GeV and 62.4 GeV~\cite{STAR_LPV1,STAR_LPV2},
showing the clear difference between the opposite sign and the same sign correlations,
qualitatively consistent with the picture of CME and LPV.
Fig.~\ref{fig:3} presents the extension of the analysis to lower beam energies at RHIC.
The STAR results are based on Au+Au samples of 57M events at 200 GeV from RHIC year 2007~\cite{Dhevan}, 
7M at 62.4 GeV (2005), 100M at 39 GeV (2010), 40M at 27 GeV (2011),
20M at 19.6 GeV (2011), 10M at 11.5 GeV (2010) and 4M at 7.7 GeV (2010).
For comparison, we also show the results for Pb+Pb collisions at 2.76 TeV~\cite{ALICE}.
A striking similarity exists between 200 GeV Au+Au and 2.76 TeV Pb+Pb,
and a smooth transition occurs from 200 GeV to lower beam energies
starting from the peripheral collisions.

\begin{figure}[htbp]
\begin{center}
 \includegraphics[width=1.0\textwidth]{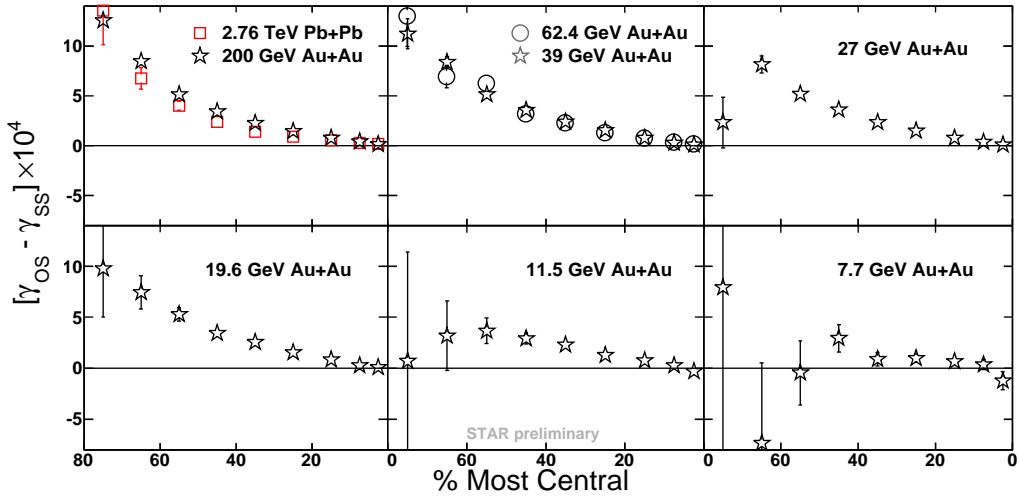}
\end{center}
\caption{(Color online) The difference between the opposite sign and the same sign correlations 
as a function of centrality for Au+Au collisions from 200 GeV to 7.7 GeV~\cite{Dhevan}.
For comparison, we also show the results for Pb+Pb collisions at 2.76 TeV~\cite{ALICE}. 
The errors are statistical only.}
\label{fig:4}
\end{figure}

Initially it was expected that the opposite sign ($\gamma_{\rm OS}$) and the same sign ($\gamma_{\rm SS}$) correlations 
would be symmetric around zero due to the charge separation induced by LPV and CME.
However, there could be common physics backgrounds in both correlations.
For example, in central collisions the strong radial flow tends to push particles to the same direction regardless 
of the charge sign, and that effect will reduce both $\gamma_{\rm OS}$ and $\gamma_{\rm SS}$ by the same amount.
In peripheral collisions, the multiplicity is smaller and the system is more influenced by momentum conservation,
which tends to increase both correlations in the same way.
Also, the statistical fluctuation of the correlator could be larger out-of-plane than in-plane
due to the geometry of the collision system, which contributes a negative background.
To reduce such mutual backgrounds, we take the difference between $\gamma_{\rm OS}$ and $\gamma_{\rm SS}$ as the signal,
shown in Fig.~\ref{fig:4}.
The signal persists almost unchanged up to 2.76 TeV and down to around 11.5 GeV, and seems to disappear at 7.7 GeV.
To be more conclusive on the transition of the signal,
more statistics are needed for collisions at 11.5 and 7.7 GeV.

\begin{figure}[htbp]
\begin{center}
 \includegraphics[width=1.0\textwidth]{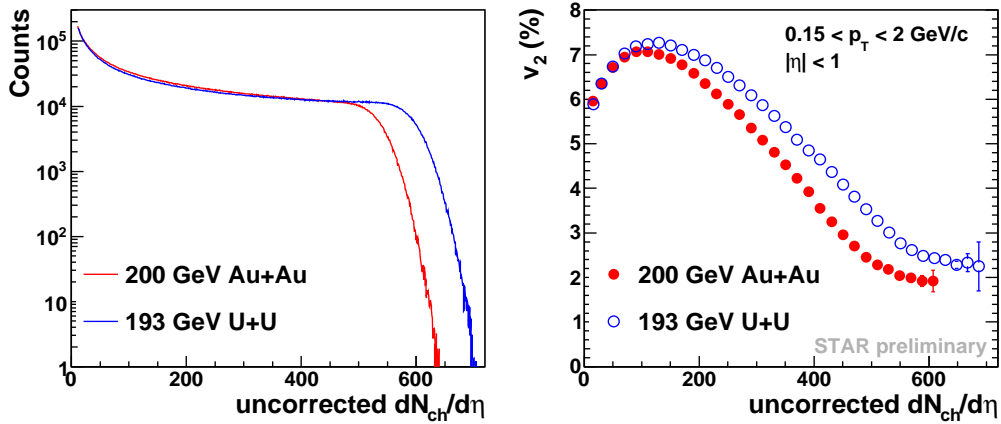}
\end{center}
\caption{(Color online) Comparison of multiplicity (left) and $p_T$-integrated $v_2$ (right) between 200 GeV Au+Au and 193 GeV U+U.
The errors are statistical only.}
\label{fig:5}
\end{figure}

One major background in $\gamma_{\rm OS} - \gamma_{\rm SS}$ comes from processes in which particles $\alpha$ and $\beta$
are products of a cluster (e.g. resonance, jet, di-jets) decay, and the cluster itself exhibits elliptic flow ($v_2$)
or different emission between in-plane and out-of-plane~\cite{Uranium}.
The corresponding contribution to the correlator can be estimated as~\cite{Resonance}
\be
\frac{N_{\rm clust/event}N_{\rm pairs/clust}}{N_{\rm pairs/event}}
\langle \cos(\phi_{\alpha}+\phi_{\beta}-2\phi_{\rm clust})\rangle_{\rm clust} \cdot v_{2,{\rm clust}},
\ee
where $\langle...\rangle_{\rm clust}$ indicates an average over pairs consisting of two daughters
from the same cluster.
In the picture of global momentum conservation~\cite{Pratt1} and/or charge conservation~\cite{Pratt2}, 
the whole system may be considered as one cluster.
To estimate this type of $v_2$-related background, we proposed to study collisions with sizeable $v_2$ and almost
no magnetic field via very central U+U collisions~\cite{Uranium}.
The U+U collisions at 193 GeV were carried out in RHIC year 2012, and the data taken by STAR contain both 
the minimum-bias trigger and a dedicated online trigger to take events with $0$-$1\%$ spectator neutrons,
so that the very central U+U collisions were selected with the magnetic field greatly suppressed.
Uranium is larger than gold, which is reflected in the multiplicity distribution in the left panel of  Fig.~\ref{fig:5}.
$v_2\{\eta {\rm~sub}\}$ vs multiplicity is shown in the right panel of Fig.~\ref{fig:5}.
The measured $v_2$ ($\sim2.5\%$) in central U+U collisions is much lower than predicted ($\sim4\%$)~\cite{Hiroshi},
which requires further understanding and investigation of the production mechanisms for multiplicity and  elliptic flow.

\begin{figure}[htbp]
\begin{center}
 \includegraphics[width=1.0\textwidth]{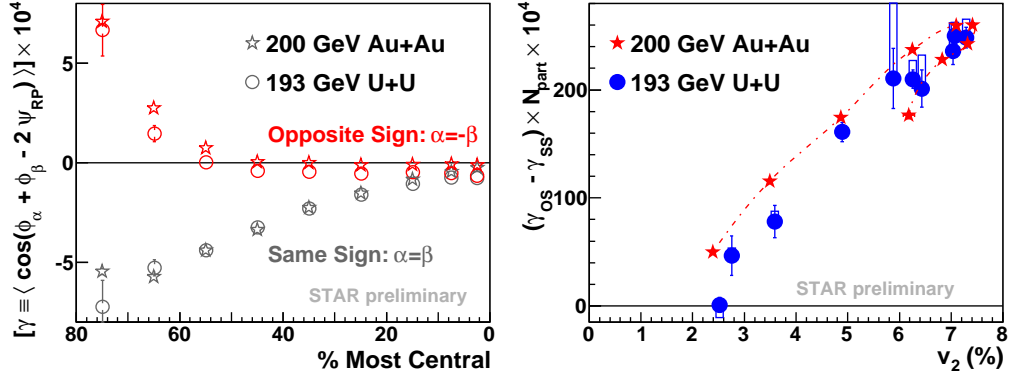}
\end{center}
\caption{(Color online) Comparison between Au+Au collisions at 200 GeV~\cite{Flow200GeV,Dhevan} and U+U collisions at 193 GeV.
(Left) $\gamma$ as a function of centrality.
(Right) $(\gamma_{\rm OS}-\gamma_{\rm SS}) \times N_{part}$ vs $v_2$.
The error bars are statistical only.
The open box represents the systematic uncertainty due to the tracking capability under the high luminosity in RHIC year 2011.}
\label{fig:6}
\end{figure}

The left panel of Fig.~\ref{fig:6} shows $\gamma$ as a function of centrality for 
U+U collisions at 193 GeV, in comparison with Au+Au collisions at 200 GeV~\cite{Flow200GeV,Dhevan}.
The opposite sign correlations in U+U are still higher than the same sign, with $\gamma_{\rm SS}$
consistent with those in Au+Au and $\gamma_{\rm OS}$ slightly lower than those in Au+Au.
To reduce the mutual background, we study $\gamma_{\rm OS} - \gamma_{\rm SS}$, and multiply it
by the number of participants, $N_{\rm part}$, to compensate for the dilution effect~\cite{STAR_LPV2}.
The right panel of Fig.~\ref{fig:6} shows the signal $(\gamma_{\rm OS} - \gamma_{\rm SS})\cdot N_{\rm part}$ 
vs $v_2$ for different centralities in 193 GeV U+U and 200 GeV Au+Au collisions.
In both U+U and Au+Au, the signal roughly increases with $v_2$, seemingly following the background trend described by Eq.(4). 
The central trigger in U+U is supposed to disentangle the background contribution from the signal,
since the magnetic field will be greatly suppressed and the measured signal will be dominated by the $v_2$-related background.
As a result, in $0$-$1\%$ most central U+U collisions the signal disappears as expected by the Chiral Magnetic Effect,
while $v_2$ is still $\sim 2.5\%$. 

\section{Summary}
The Chiral Magnetic Wave is a proposed signature of the Chiral Symmetry Restoration in the hot and dense 
nuclear matter created in heavy ion collisions at RHIC/LHC energies.
From 200 GeV to 19.6 GeV Au+Au collisions, the $v_2(A_{\pm})$ difference between $\pi^-$ and $\pi^+$
is consistentwith the qualitative expectations from the Chiral Magnetic Wave picture, 
and the slope parameter follows a centrality dependence 
qualitatively similar to the theoretical calculations of the CMW.
On the other hand, UrQMD can not reproduce the data.
Systematic investigations will be carried out to include different particle species and different $v_2$ methods.

In search of Local Parity Violation and Chiral Magnetic Effect, we studied the charge-dependent three-point correlator.
The difference between the opposite sign and the same sign correlations is present in Au+Au, Cu+Cu, U+U and Pb+Pb
collisions at RHIC and LHC, and remains almost unchanged up to 2.76 TeV and down to 11.5 GeV.
The signal, $\gamma_{\rm OS} - \gamma_{\rm SS}$, seems to disappear when the beam energy is lowered to 7.7 GeV,
or when the magnetic field is greatly suppressed as in $0$-$1\%$ most central U+U collisions,
while $v_2$ is still sizeable in both cases.
The results seem to indicate that the observed signal is not dominated by the $v_2$-related background.
A cross-check with $0$-$1\%$ most central Au+Au collisions will be carried out in future.




\begin{thebibliography}{00} 
\bibitem{Kharzeev}
D. E. Kharzeev, L. D. McLerran and H. J. Warringa, \npa {\bf 803} 227 (2008).
\bibitem{Kharzeev2}
D. Kharzeev, \plb {\bf 633} 260 (2006).
\bibitem{Kharzeev3}
D. Kharzeev and A. Zhitnitsky, \npa {\bf 797} 67 (2007).
\bibitem{Kharzeev4}
K. Fukushima, D. E. Kharzeev and H. J.Warringa, \prd {\bf 78} 074033 (2008).
\bibitem{Kharzeev5}
D. E. Kharzeev, Annals Phys. {\bf 325} 205 (2010).

\bibitem{STAR_LPV1}
B. I. Abelev {\it et al.} [STAR Collaboration], \prl {\bf 103} 251601 (2009).
\bibitem{STAR_LPV2}
B. I. Abelev {\it et al.} [STAR Collaboration], \prc {\bf 81} 054908 (2010).
\bibitem{PHENIX_LPV1}
N. N. Ajitanand, S. Esumi, R. A. Lacey [PHENIX Collaboration], in: Proc. of the RBRC Workshops, vol. 96,
2010: "P- and CP-odd effects in hot and dense matter".
\bibitem{PHENIX_LPV2}
N. N. Ajitanand, R. A. Lacey, A. Taranenko and
J. M. Alexander, arXiv:1009.5624 [nucl-ex].
\bibitem{dis1}
A. Bzdak, V. Koch and J. Liao, \prc {\bf 81} 031901
(2010); \prc {82} 054902 (2010).
\bibitem{dis2}
D. E. Kharzeev, D. T. Son, \prl {\bf 106} 062301 (2011) and references therein.
\bibitem{CSE1}
D. T. Son and A. R. Zhitnitsky, \prd {\bf 70} 074018 (2004).
\bibitem{CSE2}
M. A. Metlitski and A. R. Zhitnitsky, \prd {\bf 72} 045011 (2005).

\bibitem{CMW}
Y. Burnier, D. E. Kharzeev, J. Liao and H-U Yee,
\prl {\bf 107} 052303 (2011);
arXiv:1208.2537v1 [hep-ph].
\bibitem{CMW2}
G. M. Newman, JHEP {\bf 0601} 158 (2006).
\bibitem{Methods}
   A.M.~Poskanzer and S.A.~Voloshin,
   \prc {\bf 58}, 1671 (1998).
\bibitem{other1}
J. Dunlop, M.A. Lisa, P. Sorensen, arXiv:1107.3078.
\bibitem{other2}
J. Xu, L.-W. Chen, C. Ko, Z.-W. Lin, \prc {\bf 85} 041901 (2012).
\bibitem{other3}
J. Steinheimer, V. Koch, M. Bleicher, arXiv:1207.2791.

\bibitem{TPC-NIM}
   M. Anderson {\it et al.} [STAR Collaboration],
   Nucl. Instr. Meth. A{\bf 499}, 659 (2003).
\bibitem{Flow200GeV}
   J.~Adams {\it et al.} [STAR Collaboration],
  \prc {\bf 72}, 014904 (2005).
\bibitem{Hongwei}
   H. Ke {\it et al.} [STAR Collaboration], arXiv:1211.3216.
\bibitem{UrQMD}
   S.A. Bass {\it et al.}, Prog. Part. Nucl. Phys. {\bf 41}, 225 (1998);
   M. Bleicher {\it et al.}, J. Phys. G{\bf 25}, 1859 (1999).
\bibitem{Dhevan}
D. Gangadharan{\it et al.} [STAR Collaboration], J. Phys. G{\bf 38} 124166 (2011).
\bibitem{ALICE}
B. Abelev {\it et al.} [ALICE Collaboration], arXiv:1207.0900.

\bibitem{Lee}
T. D. Lee, \prd {\bf 8} 1226 (1973).
\bibitem{Wick}
T. D. Lee and G. C. Wick, \prd {\bf 9} 2291 (1974).
\bibitem{Morley}
P. D. Morley and I. A. Schmidt, Z. Phys. C{\bf 26} 627 (1985).
\bibitem{Dima}
D. Kharzeev, R. D. Pisarski and M. H. G. Tytgat, \prl {\bf 81} 512 (1998);
D. Kharzeev and R. D. Pisarski, \prd {\bf 61} 111901 (2000).
\bibitem{Sergei2}
S. A. Voloshin, \prc {\bf 70} 057901 (2004).

\bibitem{Uranium}
S. Voloshin, \prl {\bf 105} 172301 (2010).
\bibitem{Resonance}
S. A. Voloshin, \prc {\bf 70} 057901 (2004).

\bibitem{Pratt1}
S. Pratt, arXiv:1002.1758.
\bibitem{Pratt2}
S. Schlichting and S. Pratt, \prc {\bf 83} 014913 (2011).
\bibitem{Hiroshi}
H. Masui, B. Mohanty, N. Xu, \plb {\bf 679} 440 (2009).
\end{thebibliography}
\end{document}